# Frequency response limitation of heat flux meters

I. Naveros [a, b *], C. Ghiaus [a], D. P. Ruíz [a]

[a] Univ Lyon, CNRS, INSA-Lyon, Université Claude Bernard Lyon 1, CETHIL UMR5008, F-69621 Villeurbanne, France
[b] Department of Applied Physics, University of Granada, Campus de Fuentenueva, E-18071 Granada, Spain
[c] Energy Efficiency in buildings R&D Unit, CIEMAT, Madrid E-28040, Spain

## Abstract

Heat flux meters are used for measuring the heat flux densities going through walls, usually at steady-state. The limitations of heat flux meters under dynamic conditions are well documented in the literature; nonetheless there is a theoretical limitation which is mostly not considered and should be also taken into account. Since heat transfer is a dissipative process, it would be expected to obtain transfer functions which act as low pass filters. Nonetheless, this paper shows that the transfer functions modeling heat flow rate may become high pass filters, which is against the physical evidence. In order to show this theoretical limitation of the heat flux meters, the heat equation is transformed in different classes of models, from partial differential equations to transfer functions related to surface temperatures and heat flux density going through a wall.

# *Nomenclature*

| | | | |
|---|---|---|---|
| $b_i$ | temperature source on branch $i$, °C or K | $\mathbf{B}_S$ | input matrix in the state-space model (continuous time) |
| $C_i$ | thermal or heat capacity in node $i$, J K$^{-1}$ | $\mathbf{b}$ | vector of temperature sources on the branches |
| $C_{wall}$ | thermal capacity of the wall, J K$^{-1}$ | $\mathbf{C}$ | diagonal matrix of thermal capacities |
| $c$ | specific heat capacity, J kg$^{-1}$ K$^{-1}$ | $\mathbf{C}_S$ | output matrix in the state-space model |
| $e_i$ | temperature difference over the thermal resistance $R_i$, °C or K | $\mathbf{D}_S$ | feed through matrix in the state-space model |
| $f_i$ | heat rate source in node $i$, W | $\mathbf{f}$ | vector of heat rate sources |
| $I_s$ | solar irradiance, W m$^{-2}$ | $\mathbf{f}_C$ | vector of heat rate sources connected to nodes with a thermal capacity |
| $N$ | number of nodes | | |
| $p$ | heat sources, W m$^{-3}$ | $\mathbf{f}_0$ | vector of heat rate sources connected to nodes without a thermal capacity |
| $q_i$ | heat rate on the branch $i$, W | | |
| $Q_i$ | heat flux density inside the wall, W m$^{-2}$ | $\mathbf{G}$ | diagonal matrix of thermal conductances |
| $R_i$ | thermal resistance on the branch $i$ of the thermal network, K W$^{-1}$ | $\mathbf{H}_S$ | transfer matrix |
| $R_{wall}$ | thermal resistance of the wall, K W$^{-1}$ | $H_{ij}$ | output $i$ regarding to input $j$ component of transfer matrix |
| $R_{si}$ | indoor surface heat transfer resistance, K W$^{-1}$ | $\mathbf{H}_d$ | discrete transfer matrix |
| $R_{so}$ | outdoor surface heat transfer resistance, K W$^{-1}$ | $H_{dij}$ | output $i$ regarding to input $j$ component of discrete transfer matrix |
| $s$ | complex variable | $\mathbf{I}$ | identity matrix |
| $S$ | wall surface area, m$^2$ | $\mathbf{u}$ | input vector in the state-space model |
| $T_o, T_i$ | outdoor, indoor air temperature, °C or K | *Vector in Greek letters* | |
| *Greek letters* | | $\boldsymbol{\theta}$ | vector of temperatures |
| $\alpha$ | wall absorptivity | $\boldsymbol{\theta}_C$ | vector of temperatures in nodes with a thermal capacity |
| $\kappa$ | thermal conductivity, W m$^{-1}$K$^{-1}$ | | |
| $\theta$ | spatial temperature distribution, °C or K | $\boldsymbol{\theta}_0$ | vector of temperatures in nodes without a thermal capacity |
| $\theta_i$ | temperature of node $i$, °C or K | | |
| $\theta_{so}, \theta_{si}$ | surface temperatures outside, inside, °C or K | | |
| $\rho$ | density, kg m$^{-3}$ | | |
| $\partial/\partial t$ | differential operator in time | | |
| $\nabla \cdot$ | divergence operator | | |
| $\nabla$ | gradient operator | | |

*Vectors and matrices*

| | |
|---|---|
| $\mathbf{A}$ | incidence matrix of the thermal network |
| $\mathbf{A}^T$ | transpose of the incidence matrix |
| $\mathbf{A}_d$ | state matrix in the state-space model (discrete time) |
| $\mathbf{A}_S$ | state matrix in the state-space model (continuous time) |
| $\mathbf{B}_d$ | input matrix in the state-space model (discrete time) |

# 1   Introduction

Energy efficiency of buildings has an increased interest due to the necessity of tackling primary energy consumption in modern societies as well as societies in development (Vieites, et al., 2015). The evaluation of energy efficiency of buildings requires the measurement of their thermal performance, and particularly, thermal performance of walls which are usually studied assuming the classical heat transport theory (Carslaw & Jaeger, 1986). There are significant discrepancies between predicted and measured thermal performances of buildings and building components (de Wilde, 2014). The reason of such discrepancies may be diverse: workmanship, making tests at stationary state or at dynamic conditions, etc. (Bloem, et al., 1994; Foucquier, et al., 2013). A gap may exist between predicted and measured thermal performances, but it should be quantifiable. Therefore, efforts need to be focused on defining experimental techniques in-situ to be used so that predicted and measured thermal performances to be in agreement.

As noted, the guideline for the assessment of wall thermal performances is the classical theory of heat transfer (Carslaw & Jaeger, 1986). One usual estimation method is the heat flux meter method (Jimenez & Heras, 2005; Naveros, et al., 2012; Martin, et al., 2012; Biddulph, et al., 2014; Meng, et al., 2015), which gives a measure of the heat flux density going through a wall assuming that the heat flux meters have negligible thermal resistance in comparison with the wall thermal resistance; then, heat flux through a heat flux meter is assumed equal to heat flux through the wall on which it is plastered (EN-ISO-9869, 1994; Kaviany, 2013). Heat flux meters are useful for measuring heat flux through surfaces of homogeneous walls at steady state conditions. However, among other practical problems, heat flux through non-homogenous walls and/or under dynamic conditions is not well captured by the heat flux meters (EN-ISO-9869, 1994; Meng, et al., 2015). Anyway, in-situ measurements may be used for solving parameter identification problems, i.e. for estimating thermal resistances and capacitances of walls in order to obtain wall thermal performances of building component (Naveros, et al., 2015). For this, solar irradiance, ambient and surface temperatures are usually considered independent variables or inputs (driving functions) and heat flux density is considered a dependent variable or output (response function), **Error! Reference source not found.**. It is assumed that heat flux density going through the heat flux meters, $q_{hfm}$(W/m$^2$), which is due to the difference of heat flux meter surface temperatures, $T_{hfm1}$ and $T_{hfm2}$(°C), equals the heat flux density going through the wall. A heat flux meter gives estimations of heat flux densities by using its surface temperatures and its known thermal resistance, $R_{hfm}$(K/W).

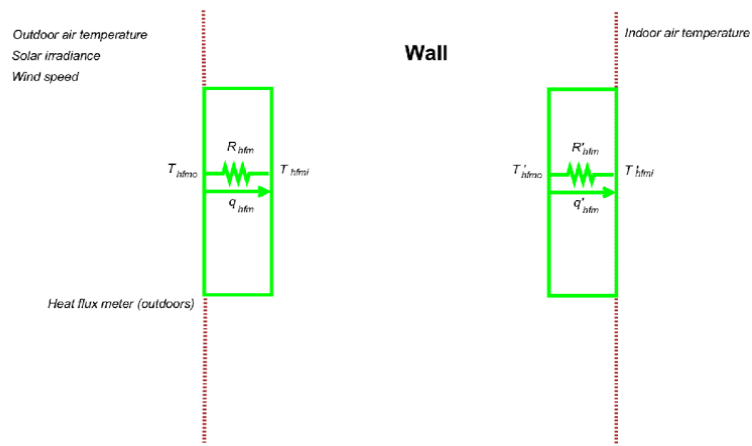

Figure 1. Heat flux meters located on the outside and inside faces of a wall

Due to experimental limitations of heat flux meters, different in-situ studies measure heat flux going through walls locating heat flux meters indoors instead of outdoors (Bloem, et al., 1994; Bacher & Madsen, 2011; Jimenez, et al., 2008; Cipriano, et al., 2016). Nonetheless, the identification problem could be solved by considering solar irradiance, ambient and surface temperatures only. In this sense, wall surface temperatures would be assumed as dependent variables or outputs for avoiding heat flux meter limitations under dynamic conditions.

This paper aims to overview well-known experimental problems related to the use of heat flux meters and to highlight a theoretical problem which appears in the use of heat flux meters, that is, transfer functions related to heat transfer may become high pass filters (Naveros & Ghiaus, 2015). Section 2 introduces heat flux meter characteristics briefly; Section 3 presents the methodology used for obtaining the transfer function representation and Section 4 shows the study of the transfer functions related to heat flux density and surface temperatures. Finally, discussion and conclusions are outlined.

## 2   Heat flux meters and error sources of measurements

Heat flux meters are sensors, **Error! Reference source not found.**, which are usually used for determining heat flux rates through walls (Naveros, et al., 2014; Biddulph, et al., 2014; Meng, et al., 2015). The heat flux sensor works by measuring the temperature difference across a reference layer with a low thermal resistance, which is known and negligible in comparison to the wall thermal resistance (Kaviany, 2013). Heat flux meters are plastered on wall surfaces using high conductivity adhesives in order to introduce a negligible surface thermal resistance.

There are several experimental problems which need to be taken into account when using heat flux meters; among them: 1) the original thermal phenomenon is distorted, which implies that the heat flux through the sensor may not correspond exactly with the heat flux through the whole wall (Meng, et al., 2015), and 2) quasi-static heat flow through walls is assumed, which is not usual for in-situ measurements.

In the first case, a solution may be to put several heat flux meters at different locations of a wall for considering their average value and assuming that the wall may be considered homogeneous using such an average value (Naveros, et al., 2012; Naveros, et al., 2014; Naveros, et al., 2015). In the second case, the solution is usually the use of heat flux meters with a time response lower enough in comparison to the dynamic changes of the driven forces which produce the heat flux through the wall. Nonetheless, heat flux meters are measuring the heat transfer through walls, which usually have large time response of minutes or hours (Naveros & Ghiaus, 2015), which implies that driven forces, such as solar irradiance, represented by a signal that contains high frequency components produce wrong measurements. This is an experimental limitation for the use of heat flux meters outdoors. Heat flux meters located outdoors could be protected from solar irradiance, but this would imply a distortion and to obtain a non-realistic measure of the heat flux through the walls which is not protected from solar irradiance. For this reason, heat flux meters are usually located indoors where they are not affected by driven forces with high frequency components as solar irradiance (Naveros & Ghiaus, 2015); in this way, a realistic measure of the heat flux through walls may be obtained.

As said above, the experimental sources of errors in the use of heat flux meters are quite important and imply limitations which are considered in the literature related to their use for wall thermal performance (Meng, et al., 2015). Nonetheless, there is also a theoretical problem related to the use of the heat flux density as output signal that is not noted. This theoretical problem has its root in the use of a classical theory, as heat conduction theory is. The problem is linked to the algebraic equation that represents heat flux since it implies that a force acts instantaneously at a distance (Roubicek & Valasek, 2002), and may give transfer functions which would act as high pass filters against the physical experience. Such a problem appears clearly in frequency domain as it will be proved in the next sections.

## 3   Methodology

The classical heat equation, which is a deterministic parabolic partial differential equation (PDE), is usually employed for modelling heat conduction in solids and is the basis of heat transfer analysis in buildings. It was introduced by Fourier in his celebrated "Théorie Analytique de la Chaleur" for studying heat diffusion from a continuous standpoint (Fourier, 1822; Narasimhan, 1999). Furthermore, Fourier developed analytical methods for solving the heat equation which are still employed (Carslaw & Jaeger, 1986; Incropera, 2006).

On the other hand, numerical methods became years later and they are mainly used for solving complex of thermal problems. Among them, finite element method has been developed significantly and numerical codes like ANSYS® Fluent or COMSOL® are widely used to solve heat conduction in combination with fluid mechanical problems. The use of these software tools in computational fluid dynamics (CFD) is mainly focused on simulation problems, and parameter identification problems are not considered usually. Nonetheless, parameter identification may be also dealt with finite element techniques and this can be noted explicitly if thermal networks and the other model structures used for modelling heat transfer are deduced from the heat equation.

The deduction of thermal networks from the heat equation supposes to follow the inverse way that Fourier followed, i.e. to discretize the heat equation for solving it. Thermal networks usually appear considering the analogy with electrical networks (Rabl, 1988; Milman & Petrick, 2000; Chen, et al., 2013), as also noted by Carslaw and Jaeger (Carslaw & Jaeger, 1986). Thus, it is important to highlight that the analogy between heat conduction and electricity started in the inverse way, i.e. Fourier's law preceded Ohm's law as noted within an historical review by Narasimhan (Narasimhan, 1999).

This section shows the deduction of a transfer matrix from the heat equation, passing through thermal networks and state-space representation, by using the finite element method (Strang, 1986; Strang, 2007).

## 3.1 Heat equation as a system of differential algebraic equations

The heat equation can be expressed as a system of linear differential and algebraic equations (DAE) using the Galerkin finite element method, as shown for systems in equilibrium by Gilbert Strang (Strang, 1986; Strang, 2007). Let us consider the heat equation for a continuous isotropic non-homogeneous medium:

$$\rho c \frac{\partial \theta}{\partial t} = -\nabla \cdot (-\kappa \nabla \theta) + p \tag{1}$$

where $\rho = \rho(x, y, z)$ is the volume density, $c = c(x, y, z)$ is the volume specific heat capacity, $\kappa = \kappa(x, y, z)$ is the thermal conductivity of the volume, $p = p(x, y, z)$ is heat rate supplied to the volume by external sources, and $\theta = \theta(x, y, z)$ is the temperature distribution in the volume.

The heat equation, Eq. (1), may be expressed in the weak form by integration of the strong form, and the finite element method may be used for solving it. In three-dimensions, the integration of the strong form of the heat equation over the volume, $V$, using a test function $v = v(x, y, z)$, is:

$$\int_V \rho c \frac{\partial \theta}{\partial t} v \mathrm{d}V = \int_V -\nabla \cdot (-\kappa \nabla \theta) v \mathrm{d}V + \int_V p v \mathrm{d}V \tag{2}$$

The weak form appears once the first term in the right-hand side of Eq. (**2**) is integrated by parts. Firstly, the divergence and gradient operators may be developed considering Cartesian coordinates:

$$\int_V \rho c \frac{\partial \theta}{\partial t} v \mathrm{d}V = \int_V \left( \frac{\partial}{\partial x}\left(\kappa \frac{\partial \theta}{\partial \mathrm{x}}\right) + \frac{\partial}{\partial y}\left(\kappa \frac{\partial \theta}{\partial \mathrm{y}}\right) + \frac{\partial}{\partial z}\left(\kappa \frac{\partial \theta}{\partial \mathrm{z}}\right) \right) v \mathrm{d}V + \int_V p v \mathrm{d}V \tag{3}$$

This integral expression becomes a system of differential algebraic equations (DAE), and may be represented graphically as a thermal network. Next, the part corresponding to the $x$-coordinate is used for detailing how obtaining the matrix form. The extension to three dimensions may be done considering a test function $v = v(x) + v(y) + v(z)$.

For one dimension, considering the total interval $x \in [0,1]$, we obtain:

$$\int_S \mathrm{d}y\mathrm{d}z \int_0^1 \rho c \frac{\partial \theta}{\partial t} v \mathrm{d}x = \int_S \mathrm{d}y\mathrm{d}z \int_0^1 \frac{\partial}{\partial x}\left(\kappa \frac{\partial \theta}{\partial \mathrm{x}}\right) v \mathrm{d}x + \int_S \mathrm{d}y\mathrm{d}z \int_0^1 p v \mathrm{d}x \tag{4}$$

Moreover, by considering the integration into a cube of volume: $V=\int_V dx\,dy\,dz = h^3$, the surface perpendicular to the $x$-coordinate will be, $S = \int_S dy\,dz = h^2$, and Eq. (4) becomes:

$$h^2 \int_0^1 \rho\, c \frac{\partial \theta}{\partial t} v\, dx = h^2 \int_0^1 \frac{\partial}{\partial x}\left(\kappa \frac{\partial \theta}{\partial x}\right) v\, dx + h^2 \int_0^1 p\, v\, dx \qquad (5)$$

Following, Eq. (5) will be put in matrix notation by using the finite element Galerkin method (Strang, 2007). For this purpose, it is used a continuous piecewise linear approximation for building the trial functions $\varphi_i(x)$, Figure 2. Let $I_h: 0 = x_1 < x_2 < \cdots < x_N = 1$ be a partition of interval $I = (0,1)$ into subintervals $I_j = (x_j, x_{j-1})$ of length $h_j = x_j - x_{j-1}$; we will suppose for the sake of simplicity $h_j = h$. Let $V_h$ denote the set of continuous piecewise linear functions on $I_h$ that are one at $x = 0$ and $x = 1$. $V_h$ is a finite dimensional vector space, $dim V_h = N$, with a basis consisting of hat functions $\{\varphi_i\}_{i=1}^N$ :

$$\varphi_i(x) = \begin{cases} 0, & if\ x \notin [x_{i-1}, x_{i+1}], \\ \dfrac{x - x_{i-1}}{x_i - x_{i-1}}, & if\ x \in [x_{i-1}, x_i], \\ \dfrac{x_{i+1} - x}{x_{i+1} - x_i}, & if\ x \in [x_i, x_{i+1}]; \end{cases} \qquad (6)$$

Using such trial functions, the temperature function may be defined as:

$$\theta(x) = \sum_{i=1}^N \theta_i\, \varphi_i(x) \qquad (7)$$

and the test function may be defined as a set of $N$ functions at each space interval by:

$$v_j = \varphi_j(x)\ \ j = 2, \ldots, N-1;\ v_1 = 0;\ v_N = 0 \qquad (8)$$

These test functions will coincide with hat functions except at interval extremes ($v_1 \neq \varphi_1, v_N \neq \varphi_N$).

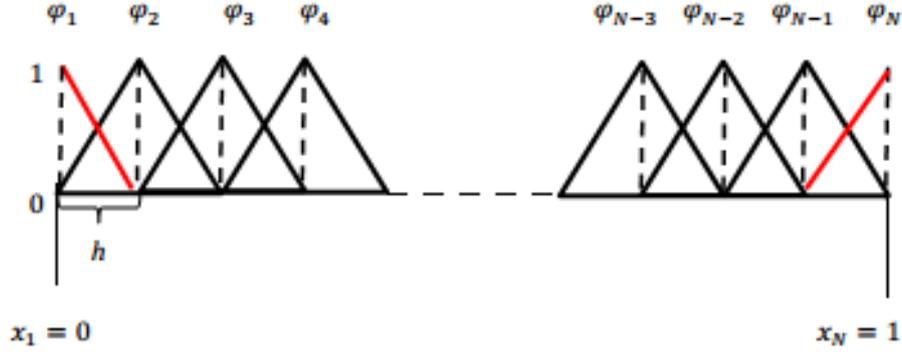

Figure 2. Trial hat functions and semi-hat functions at extremes

The function of thermal conductivities may be defined as being constant on intervals:

$$\kappa(x) = \kappa_i \quad for\ x \in\ ]x_{i-1}, x_i[, \quad i = 1, \ldots, N-1 \qquad (9)$$

It needs to be noted that thermal conductivity does not need to be a continuous function as are the heat flux and the temperature distribution. At interval junctions, the thermal conductivity may change without continuity since the space may be non-homogenous; for instance, when there is a change of material (Carslaw & Jaeger, 1986; Strang, 2007).

Next, we consider the integration by parts of the first term in the right-hand side of Eq. (5):

$$h^2 \int_0^1 \frac{d}{dx}\left(\kappa \frac{d\theta}{dx}\right) v\, dx = -h^2 \int_0^1 \kappa \frac{d\theta}{dx} \frac{dv}{dx} dx + h^2 \kappa \frac{d\theta}{dx} v \Big|_0^1 \qquad (10)$$

The second term in the right-hand side becomes zero by choosing the test function $v(0, y, z) = 0$ and $v(1, y, z) = 0$; the first term in the right-hand side of Eq. (10) may be written as $N$ expressions ($j = 1, \ldots, N$) and each expression can be expanded in $N$ terms ($i = 1, \ldots, N$) considering the previously proposed test and trial functions ($v_j$, $\varphi_i$):

$$\left.\begin{array}{c} -h^2\left(\dfrac{d\varphi_1}{dx}\theta_1 + \cdots + \dfrac{d\varphi_N}{dx}\theta_N\right)\left(\dfrac{dv_1}{dx}\right)\int_0^1 \kappa dx \\ \vdots \\ -h^2\left(\dfrac{d\varphi_1}{dx}\theta_1 + \cdots + \dfrac{d\varphi_N}{dx}\theta_N\right)\left(\dfrac{dv_N}{dx}\right)\int_0^1 \kappa dx \end{array}\right\} N\ test\ functions\ (v_j) \qquad (11)$$

with the overbrace: $N\ trial\ functions\ (\varphi_i)$

This enables to use of matrix notation for rewriting expression (11) as $\mathbf{K\theta}_\varphi$, where $\mathbf{\theta}_\varphi = (\theta_1 \ldots \theta_n)^T$ is the vector of temperatures representing state variables and temperature sources and $\mathbf{K}$ is the stiffness matrix, which is a sparse tridiagonal matrix ($N \times N$) with elements different to zero only on the diagonal and on the adjacent diagonals. Furthermore, this stiffness matrix will be symmetric and positive definite, hence invertible.

In practice, for obtaining the stiffness matrix the test and trial functions, which are piece-wise functions, may be arranged as vectors and matrices. The total interval [0,1] may be split into $N - 1$ subintervals, which may be supposed equal spaced with a length $h$, and each subinterval will represent a vector component. For instance, the derivatives of a set of trial functions, Eq. (6), considering $N = 5$, may be represented as five independent column vectors and arranged as a matrix, Figure 3-Figure 7 and Eq. (12)-(18).

The derivative of the first trial function ($\varphi_1'$), Figure 3, may be defined for each interval as:

$$\boldsymbol{\varphi}'_1 = \frac{1}{h}(-1\ \ 0\ \ 0\ \ 0)^T \qquad (12)$$

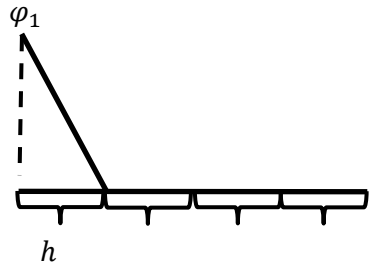

Figure 3. First trial function (semi-hat function) using four intervals of length $h$.

The derivative of the second trial function ($\varphi_2'$) considered in Figure 4 may be defined for each interval as 1/h, -1/h, 0 and 0 respectively:

$$\boldsymbol{\varphi'}_2 = \frac{1}{h}\begin{pmatrix}1 & -1 & 0 & 0\end{pmatrix}^T \tag{13}$$

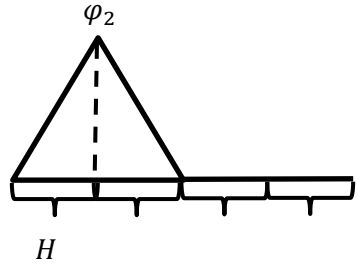

Figure 4. Second trial function (hat function) using four intervals of length *h*.

The derivative of the third trial function ($\varphi'_3$) considered in Figure 5 may be defined for each interval as 0, 1/h, -1/h and 0 respectively:

$$\boldsymbol{\varphi'}_3 = \frac{1}{h}\begin{pmatrix}0 & 1 & -1 & 0\end{pmatrix}^T \tag{14}$$

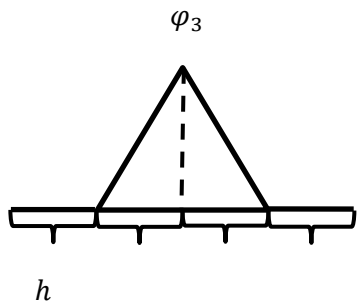

Figure 5. Third trial function (hat function) using four intervals of length *h*

The derivative of the fourth trial function ($\varphi'_4$) considered in Figure 6 may be defined for each interval as 0, 0, 1/h and -1/h respectively:

$$\boldsymbol{\varphi'}_4 = \frac{1}{h}\begin{pmatrix}0 & 0 & 1 & -1\end{pmatrix}^T \tag{15}$$

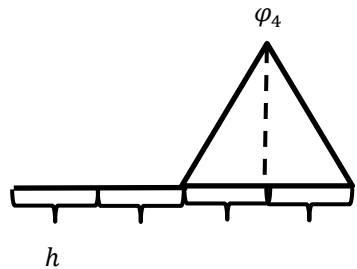

Figure 6. Fourth trial function (hat function) using four intervals of length *h*

The derivative of the fifth trial function ($\varphi_5'$) considered in Figure 7 may be defined for each interval as 0, 0, 0 and 1/h respectively:

$$\boldsymbol{\varphi}'_5 = \frac{1}{h}\begin{pmatrix} 0 & 0 & 0 & 1 \end{pmatrix}^T \tag{16}$$

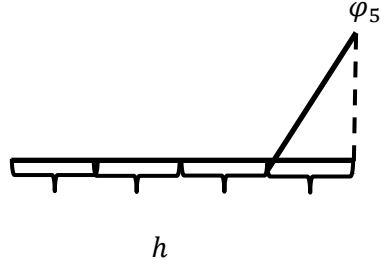

Figure 7. Fifth trial function (semi-hat function) using four intervals of length $h$

The trial functions may be arranged altogether as column vectors to form a difference matrix, $\mathbf{A}_\varphi$:

$$\mathbf{A}_\varphi = \frac{1}{h}\begin{pmatrix} \boldsymbol{\varphi}'_1 & \boldsymbol{\varphi}'_2 & \boldsymbol{\varphi}'_3 & \boldsymbol{\varphi}'_4 & \boldsymbol{\varphi}'_5 \\ -1 & 1 & 0 & 0 & 0 \\ 0 & -1 & 1 & 0 & 0 \\ 0 & 0 & -1 & 1 & 0 \\ 0 & 0 & 0 & -1 & 1 \end{pmatrix} \tag{17}$$

On the other hand, the test functions are equal to the trial functions except at interval extremes, where they may be imposed to be zero for cancelling the second term on the right-hand side of Eq. (10). This second term is related to the boundary conditions used for solving the problem and it needs to be cancelled for obtaining an invertible stiffness matrix, which is similar to the use of boundary conditions after an analytical integration for obtaining the integration constants. The difference matrix, $\mathbf{A}_v$, obtained from test functions is:

$$\mathbf{A}_v = \frac{1}{h}\begin{pmatrix} v'_1 & v'_2 & v'_3 & v'_4 & v'_5 \\ 0 & 1 & 0 & 0 & 0 \\ 0 & -1 & 1 & 0 & 0 \\ 0 & 0 & -1 & 1 & 0 \\ 0 & 0 & 0 & -1 & 0 \end{pmatrix} \tag{18}$$

Thermal conductivities may be arranged in a diagonal matrix, $\mathbf{\Lambda}$, considering they can be different for each interval ($N - 1 = 4$) of length $h$:

$$\mathbf{\Lambda} = \begin{pmatrix} \kappa_1 & 0 & 0 & 0 \\ 0 & \kappa_2 & 0 & 0 \\ 0 & 0 & \kappa_3 & 0 \\ 0 & 0 & 0 & \kappa_4 \end{pmatrix} \tag{19}$$

In this way, expression (11) may be written using matrix notation as:

$$-h^2 \mathbf{A}_v^T h \mathbf{\Lambda} \mathbf{A}_\varphi \boldsymbol{\theta}_\varphi \tag{20}$$

where the term $h\mathbf{\Lambda}$ comes from the integral given in expression (11), which is integrated interval by interval:

$$\int_0^1 \kappa \, dx = \int_0^h \kappa_1 \, dx + \int_h^{2h} \kappa_2 \, dx + \int_{2h}^{3h} \kappa_3 \, dx + \int_{3h}^1 \kappa_4 \, dx \tag{21}$$

With this configuration ($N = 5$), we have five trial functions, $\varphi_i$, and five test functions, $v_j$, but only three test functions ($v_2, v_3, v_4$) are different from zero. Hence three equations obtained from expression (20) are only different from zero:

$$-\begin{pmatrix} 0 & 0 & 0 & 0 \\ 1 & -1 & 0 & 0 \\ 0 & 1 & -1 & 0 \\ 0 & 0 & 1 & -1 \\ 0 & 0 & 0 & 0 \end{pmatrix} h \begin{pmatrix} \kappa_1 & 0 & 0 & 0 \\ 0 & \kappa_2 & 0 & 0 \\ 0 & 0 & \kappa_3 & 0 \\ 0 & 0 & 0 & \kappa_4 \end{pmatrix} \begin{pmatrix} -1 & 1 & 0 & 0 & 0 \\ 0 & -1 & 1 & 0 & 0 \\ 0 & 0 & -1 & 1 & 0 \\ 0 & 0 & 0 & -1 & 1 \end{pmatrix} \begin{pmatrix} \theta_{\varphi 1} \\ \theta_{\varphi 2} \\ \theta_{\varphi 3} \\ \theta_{\varphi 4} \\ \theta_{\varphi 5} \end{pmatrix} = h \begin{pmatrix} 0 \\ \kappa_1(\theta_{\varphi 1} - \theta_{\varphi 2}) + \kappa_2(\theta_{\varphi 3} - \theta_{\varphi 2}) \\ \kappa_2(\theta_{\varphi 2} - \theta_{\varphi 3}) + \kappa_3(\theta_{\varphi 4} - \theta_{\varphi 3}) \\ \kappa_3(\theta_{\varphi 3} - \theta_{\varphi 4}) - \kappa_4(\theta_{\varphi 5} - \theta_{\varphi 4}) \\ 0 \end{pmatrix} \tag{22}$$

The result obtained in Eq. (22) may be rearranged as:

$$-\mathbf{A}^T \mathbf{G} \mathbf{A} \boldsymbol{\theta} + \mathbf{A}^T \mathbf{G} \mathbf{b} \tag{23}$$

where the vector of temperatures, $\boldsymbol{\theta}_\varphi = (\theta_{\varphi 1} \; \theta_{\varphi 2} \; \theta_{\varphi 3} \; \theta_{\varphi 4} \; \theta_{\varphi 5})^T$, is split in a vector of temperatures representing the state-variables, $\boldsymbol{\theta} = (\theta_1 \; \theta_2 \; \theta_3)^T$, and in a vector of temperature sources (boundary conditions), $\mathbf{b} = (b_1 \; 0 \; 0 \; b_4)^T$. For this, the next change of variables is considered: $\theta_{\varphi 1} \equiv b_1$, $\theta_{\varphi 2} \equiv \theta_1$, $\theta_{\varphi 3} \equiv \theta_2$ and $\theta_{\varphi 4} \equiv \theta_3$ and $\theta_{\varphi 5} \equiv -b_4$. $\mathbf{A}$ is a difference matrix where all the columns are independent, which guarantees that the stiffness matrix ($\mathbf{K} \equiv -\mathbf{A}^T \mathbf{G} \mathbf{A}$) is invertible:

$$\mathbf{A} = \begin{pmatrix} 1 & 0 & 0 \\ -1 & 1 & 0 \\ 0 & -1 & 1 \\ 0 & 0 & -1 \end{pmatrix} \tag{24}$$

and **G** is a matrix of thermal conductances, which is obtained by multiplying the length of each interval, $h$, by the matrix of thermal conductivities, **Λ**:

$$\mathbf{G} = h\mathbf{\Lambda} = \begin{pmatrix} R_1^{-1} & 0 & 0 & 0 \\ 0 & R_2^{-1} & 0 & 0 \\ 0 & 0 & R_3^{-1} & 0 \\ 0 & 0 & 0 & R_4^{-1} \end{pmatrix} \tag{25}$$

Next, the terms which refer the heat rate sources and the energy accumulation in Eq. (5) may be put altogether as:

$$h^2 \int_0^1 \left( \rho c \frac{\partial \theta}{\partial t} - p \right) v \, \mathrm{d}x \tag{26}$$

The expression (26) may be also written in matrix notation. For instance, by making the integration using the Lagrange interpolation method:

$$\int_0^1 g(x) v_i(x) \, \mathrm{d}x \approx g(x_i) \int_0^1 v_i(x) \, \mathrm{d}x \tag{27}$$

where in our case for $g = h^2(\rho c \dot{\theta} - p)$.

Eq. (27) may be split in five terms corresponding to five test functions ($N = 5$), but only three test functions will be different from zero as in Eq. (22). This allows to define the matrix of thermal capacities, **C**, as:

$$\mathbf{C} = \begin{pmatrix} C_1 & 0 & 0 \\ 0 & C_2 & 0 \\ 0 & 0 & C_3 \end{pmatrix} \tag{28}$$

where $C_i = \rho_i c_i h^3$; the vector of heat rate sources is defined as:

$$\mathbf{f} = (f_1 \ f_2 \ f_3)^T \tag{29}$$

where $f = p_i h^3$.

The resulting matrix expression is:

$$\mathbf{C\dot{\theta}} - \mathbf{f} \tag{30}$$

where $\mathbf{\dot{\theta}}$ is the time derivative of the vector of temperatures at nodes, $\mathbf{\theta}$.

Next, by combining Eq. (23) and (30), a system of differential algebraic equations (DAE), i.e. a thermal network, Figure 8, may be deduced from the heat equation. In this way, a thermal model may be expressed as a system of differential algebraic equations (DAE), supposed linear and time invariant (Ghiaus, 2013):

$$\mathbf{C\dot{\theta}} = -\mathbf{A}^T\mathbf{GA}\mathbf{\theta} + \mathbf{A}^T\mathbf{Gb} + \mathbf{f} \tag{31}$$

Finally, a system of DAE can be represented as a thermal network, Figure 8, as shown by Ghiaus and Naveros et al. (Ghiaus, 2013; Naveros & Ghiaus, 2015; Naveros, et al., 2015).

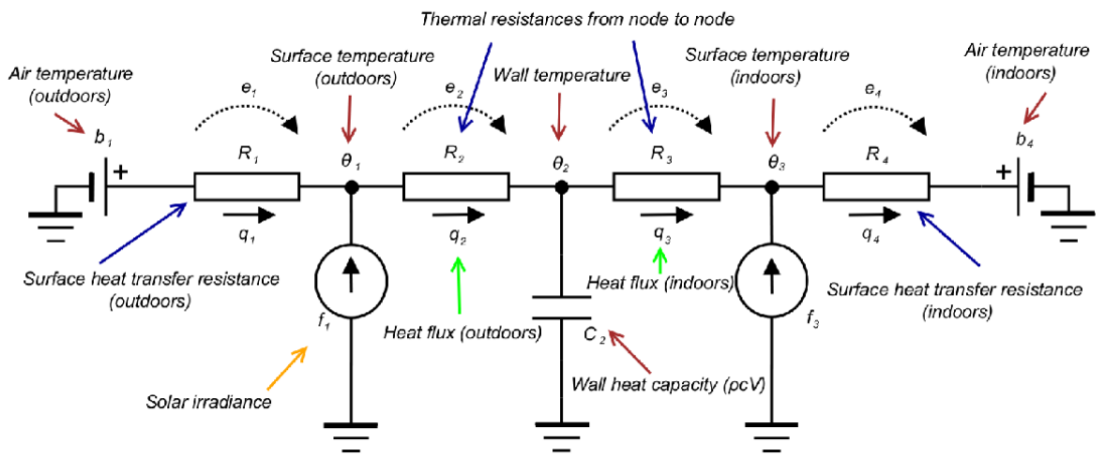

Figure 8. A first order thermal network representing a wall from which a system of differential and algebraic equations may be obtained

In Figure 8, $b_1$ and $b_4$ are temperature sources which act as boundary conditions, $f_1$ and $f_3$ are heat rate sources which act as boundary conditions, $\theta_1, \theta_2,$ and $\theta_3$ are temperatures at nodes, $R_1, R_2, R_3$ and $R_4$ are thermal resistances in branches from node to node, $q_1, q_2, q_3,$ and $q_4$ are heat rate fluxes over branches, and $C_2$ is the thermal capacity at node $\theta_2$. It is important to note that the thermal network is not necessarily one-dimensional, therefore it should be understood as a simplified graphical representation of the physical system and the system of DAE. Figure 9 shows that each node may represent a volume rather than a surface or a single point.

The first order model is presented in Figure 9 as a balance energy scheme for showing its connection with the thermal network, Figure 8. A priori, lateral fluxes could be considered negligible if lateral surfaces are much smaller than the front surfaces and the wall is built with high lateral insulation contacts (Strachan & Baker, 2008).

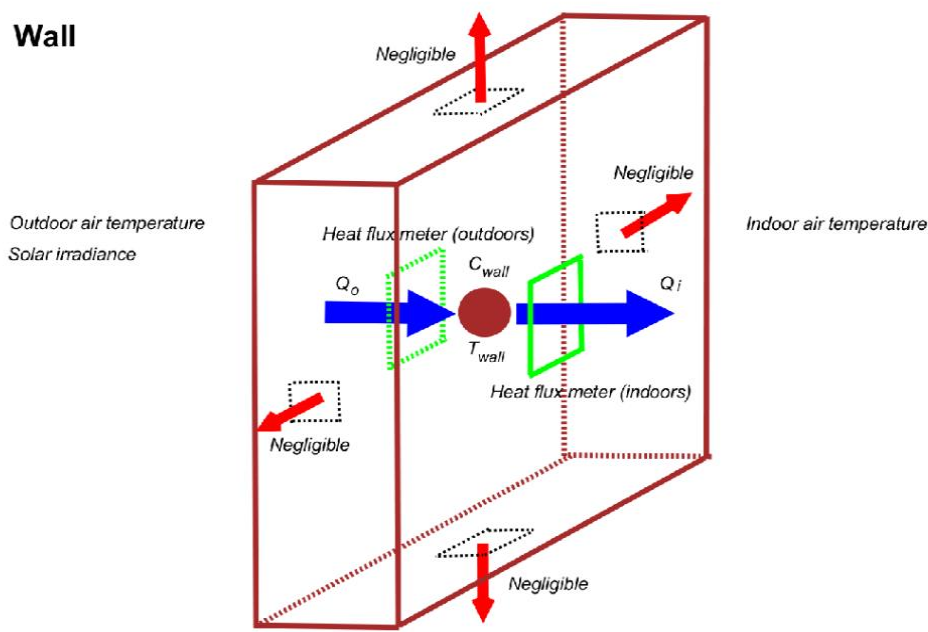

Figure 9. Wall as a single volume, energy balance scheme

In this particular example, the thermal network and the energy balance scheme, Figure 8 and Figure 9, consider a first order model for representing a wall as a single volume where:

- $\theta_1$ represents the temperature of the wall outside surface, that is, the node represents a surface;
- $\theta_2$ represents the temperature of the wall considered as a single volume ($T_{wall}$), that is, the node represents a volume;

- $\theta_3$ represents the temperature of the wall inside surface, that is, the node represents a surface;
- $R_1$ represents the outdoor surface heat transfer resistance from the outside source of temperature to the node representing the outside surface;
- $R_2, R_3$ represent the thermal resistance of heat coming in/out the node representing the wall as a single volume from nodes representing wall surfaces;
- $R_4$ represents the indoor surface heat transfer resistance from the inside source of temperature to the node representing the inside surface;
- $q_2 \equiv Q_o$ represents the heat flux from the outside surface to the single volume, which could be measured by a heat flux meter located on the outside wall surface;
- $q_3 \equiv Q_i$ represents the heat flux from the single volume to the inside surface, which could be measured by a heat flux meter located on the inside wall surface; and
- $C_2 \equiv C_{wall}$ represents the thermal capacitance of the wall, which is given by the density, the specific heat capacity and the volume, that is, $C_{wall} = \rho c V$.

In a first order model, the whole wall is assumed to have a single thermal capacitance and a single temperature; this also supposes an a priori simplification of the problem which should be checked a posteriori in practical applications. The right selection of the order of the model may be done by frequency analysis of measurements (Naveros & Ghiaus, 2015). Both a priori and a posteriori assumptions should be also checked by a cross validation of the measured outputs compared to the simulated outputs.

## 3.2 State-space representation of the thermal model

The system of differential algebraic equations (DAE) can be put in state-space representation. The system of DAE, Eq. (31), can be expressed as:

$$\mathbf{C}\dot{\boldsymbol{\theta}} = \mathbf{K}\boldsymbol{\theta} + \mathbf{K}_b \mathbf{b} + \mathbf{f} \tag{32}$$

where, by notation $\mathbf{K} = -\mathbf{A}^T \mathbf{G} \mathbf{A}$ and $\mathbf{K}_b = \mathbf{A}^T \mathbf{G}$. By considering the temperature nodes without heat capacity and those with heat capacity, the differential equations can be separated from algebraic equations and Eq. (32) expressed in blocks as:

$$\begin{bmatrix} \mathbf{0} & \mathbf{0} \\ \mathbf{0} & \mathbf{C}_C \end{bmatrix} \begin{bmatrix} \dot{\boldsymbol{\theta}}_0 \\ \dot{\boldsymbol{\theta}}_C \end{bmatrix} = \begin{bmatrix} \mathbf{K}_{11} & \mathbf{K}_{12} \\ \mathbf{K}_{21} & \mathbf{K}_{22} \end{bmatrix} \begin{bmatrix} \boldsymbol{\theta}_0 \\ \boldsymbol{\theta}_C \end{bmatrix} + \begin{bmatrix} \mathbf{K}_{b1} \\ \mathbf{K}_{b2} \end{bmatrix} \mathbf{b} + \begin{bmatrix} \mathbf{I}_{11} & \mathbf{0} \\ \mathbf{0} & \mathbf{I}_{22} \end{bmatrix} \begin{bmatrix} \mathbf{f}_0 \\ \mathbf{f}_C \end{bmatrix} \quad (33)$$

By eliminating the algebraic equations from DAEs represented by Eq. (33), the state-space representation can be obtained:

$$\dot{\boldsymbol{\theta}}_C = \mathbf{A}_S \boldsymbol{\theta}_C + \mathbf{B}_S \mathbf{u} \quad (34)$$

Once Eq. (34) is solved for the state variables, the other outputs can be obtained by using an algebraic equation:

$$\boldsymbol{\theta}_0 = \mathbf{C}_S \boldsymbol{\theta}_C + \mathbf{D}_S \mathbf{u} \quad (35)$$

### 3.2.1 Differential equations

The relation between blocks and the state matrix, $\mathbf{A}_S$, is:

$$\mathbf{A}_S = \mathbf{C}_C^{-1}(-\mathbf{K}_{21}\mathbf{K}_{11}^{-1}\mathbf{K}_{12} + \mathbf{K}_{22}) \quad (36)$$

and the relation between blocks and the input matrix, $\mathbf{B}_S$, is:

$$\mathbf{B}_S = \mathbf{C}_C^{-1}[-\mathbf{K}_{21}\mathbf{K}_{11}^{-1}\mathbf{K}_{b1} + \mathbf{K}_{b2} \quad -\mathbf{K}_{21}\mathbf{K}_{11}^{-1} \quad \mathbf{I}_{22}] \quad (37)$$

The input vector is given by:

$$\mathbf{u} = [\mathbf{b} \quad \mathbf{f}_0 \quad \mathbf{f}_C]^T \quad (38)$$

### 3.2.2 Algebraic equations

From Eq. (33), we can also obtain the system of algebraic equations which completes the state-space model:

$$\begin{aligned}\boldsymbol{\theta}_0 &= -\mathbf{K}_{11}^{-1}(\mathbf{K}_{12}\boldsymbol{\theta}_C + \mathbf{K}_{b1}\mathbf{b} + \mathbf{I}_{11}\mathbf{f}_0) \\ &= -\mathbf{K}_{11}^{-1}\left(\mathbf{K}_{12}\boldsymbol{\theta}_C + [\mathbf{K}_{b1} \quad \mathbf{I}_{11} \quad \mathbf{0}]\begin{bmatrix}\mathbf{b}\\ \mathbf{f}_0 \\ \mathbf{f}_C\end{bmatrix}\right)\end{aligned} \quad (39)$$

The relation between blocks and the output matrix, $\mathbf{C}_S$, is:

$$\mathbf{C}_S = -\mathbf{K}_{11}^{-1}\mathbf{K}_{12} \quad (40)$$

and the relation between blocks and the feed through matrix, $\mathbf{D}_S$, is

$$\mathbf{D}_S = -\mathbf{K}_{11}^{-1}[\mathbf{K}_{b1} \quad \mathbf{I}_{11} \quad \mathbf{0}] \quad (41)$$

## 3.3 Transfer function representation of thermal model: Laplace transform

The relation between the inputs, $\mathbf{u}$, and the outputs, $\boldsymbol{\theta}_0$, can be expressed as a set of transfer functions or a transfer matrix. Applying Laplace transform, for zero initial conditions, Eq. (34)-(35) can be represented as:

$$s\boldsymbol{\theta}_C = \mathbf{A}_S\boldsymbol{\theta}_C + \mathbf{B}_S\mathbf{u} \quad (42)$$

$$\boldsymbol{\theta}_0 = \mathbf{C}_S\boldsymbol{\theta}_C + \mathbf{D}_S\mathbf{u} \quad (43)$$

where $s = \sigma + j\omega$ is the complex variable. From Eq. (42)-(43), it is possible to obtain:

$$\boldsymbol{\theta}_0 = (\mathbf{C}_S(s\mathbf{I} - \mathbf{A}_S)^{-1}\mathbf{B}_S + \mathbf{D}_S)\mathbf{u} \quad (44)$$

The transfer matrix is:

$$\mathbf{H}_S = \mathbf{C}_S(s\mathbf{I} - \mathbf{A}_S)^{-1}\mathbf{B}_S + \mathbf{D}_S \quad (45)$$

Eq. (45) is composed of transfer functions relating each output, of the output vector $\boldsymbol{\theta}_0$, with a particular input from the input vector $\mathbf{u}$. For a wall, Figure 8, the transfer function can be written as:

$$\mathbf{H}_S = \begin{bmatrix} \mathbf{H}_1 & \mathbf{H}_2 \end{bmatrix}^T = \begin{bmatrix} H_{11} & H_{12} & H_{13} \\ H_{21} & H_{22} & H_{23} \end{bmatrix} \tag{46}$$

If the surface temperatures, which are nodes with negligible heat capacity, are used as outputs, then:

$$\mathbf{H}_1 = \begin{bmatrix} \frac{\theta_{so}}{T_o} & \frac{\theta_{so}}{T_i} & \frac{\theta_{so}}{I_s} \end{bmatrix}; \mathbf{H}_2 = \begin{bmatrix} \frac{\theta_{si}}{T_o} & \frac{\theta_{si}}{T_i} & \frac{\theta_{si}}{I_s} \end{bmatrix}.$$

If the heat flux densities are used as outputs, then:

$$\mathbf{H}_1 = \begin{bmatrix} \frac{Q_o}{T_o} & \frac{Q_o}{T_i} & \frac{Q_o}{I_s} \end{bmatrix}; \mathbf{H}_2 = \begin{bmatrix} \frac{Q_i}{T_o} & \frac{Q_i}{T_i} & \frac{Q_i}{I_s} \end{bmatrix}.$$

Then, the transfer functions can be obtained from state-space matrices to be used for a frequency domain analysis of heat flux density and surface temperatures considered as outputs of the system.

## 4 Frequency study of not strictly proper transfer functions

### 4.1 State-space for a first order model

#### 4.1.1 Differential equation

From the wall represented by the thermal network of Figure 8 (noting $R_1 \equiv R_{so}; R_4 \equiv R_{si}$), we obtain:

$$\mathbf{A}_S = \frac{1}{C_2}\left(-\frac{R_{so}^{-1}R_2^{-1}}{R_{so}^{-1}+R_2^{-1}} - \frac{R_3^{-1}R_{si}^{-1}}{R_3^{-1}+R_{si}^{-1}}\right), \text{ which is the state matrix;}$$

$$\mathbf{B}_S = \frac{1}{C_2}\begin{bmatrix} \frac{R_{so}^{-1}R_2^{-1}}{R_{so}^{-1}+R_2^{-1}} & \frac{-R_2^{-2}}{R_{so}^{-1}+R_2^{-1}} + R_2^{-1} & \frac{R_3^{-2}}{R_3^{-1}+R_{si}^{-1}} - R_3^{-1} & -\frac{R_3^{-1}R_{si}^{-1}}{R_3^{-1}+R_{si}^{-1}} & \frac{R_2^{-1}}{R_{so}^{-1}+R_2^{-1}} & \frac{R_3^{-1}}{R_3^{-1}+R_{si}^{-1}} & 1 \end{bmatrix},$$

which is the input matrix; and

$\mathbf{u} = [T_o \ 0 \ 0 \ -T_i \ \alpha S I_s \ 0 \ 0]^T$, which is the input vector.

The state vector is $\boldsymbol{\theta}_C = \theta_2$. By noting $\theta_2 \equiv \theta$ the next differential equation is obtained:

$$\dot{\theta} = \frac{1}{C_2} \left( \frac{R_{so}^{-1} R_2^{-1}}{R_{so}^{-1} + R_2^{-1}} (T_o - \theta) + \frac{R_3^{-1} R_{si}^{-1}}{R_3^{-1} + R_{si}^{-1}} (T_i - \theta) + \frac{R_2^{-1}}{R_{so}^{-1} + R_2^{-1}} \alpha SI_s \right) \quad (47)$$

### 4.1.2 Algebraic equations

On the other hand, Figure 8, we obtain the algebraic equations:

$$\mathbf{C}_S = \begin{bmatrix} \frac{R_2^{-1}}{R_{so}^{-1} + R_2^{-1}} \\ \frac{R_3^{-1}}{R_3^{-1} + R_{si}^{-1}} \end{bmatrix}, \text{ which is the output matrix, and}$$

$$\mathbf{D}_S = \begin{bmatrix} \frac{R_{so}^{-1}}{R_{so}^{-1} + R_2^{-1}} & \frac{R_2^{-1}}{R_{so}^{-1} + R_2^{-1}} & 0 & 0 & \frac{1}{R_{so}^{-1} + R_2^{-1}} & 0 & 0 \\ 0 & 0 & \frac{R_3^{-1}}{R_3^{-1} + R_{si}^{-1}} & \frac{-R_{si}^{-1}}{R_3^{-1} + R_{si}^{-1}} & 0 & \frac{1}{R_3^{-1} + R_{si}^{-1}} & 0 \end{bmatrix}, \text{ which is the feed}$$

through matrix.

We also obtain two measurement equations for surface temperatures:

$$\theta_{so} = \frac{R_2^{-1}}{R_{so}^{-1} + R_2^{-1}} \theta + \frac{R_{so}^{-1}}{R_{so}^{-1} + R_2^{-1}} T_o + \frac{1}{R_{so}^{-1} + R_2^{-1}} \alpha SI_s \quad (48)$$

$$\theta_{si} = \frac{R_3^{-1}}{R_3^{-1} + R_{si}^{-1}} \theta + \frac{R_{si}^{-1}}{R_3^{-1} + R_{si}^{-1}} T_i \quad (49)$$

where $\boldsymbol{\theta_{so} \equiv \theta_1}$, $\boldsymbol{\theta_{si} \equiv \theta_3}$ and $\boldsymbol{f_3 \equiv 0}$.

### 4.1.3 Extra algebraic equation

As in previous studies (Naveros, et al., 2014), measurement equations related to the use of a heat flux meter located in the inner or outer face of the wall may be also considered, Figure 8:

$$Q_i = \frac{1}{S} \frac{R_3^{-1} R_{si}^{-1}}{R_3^{-1} + R_{si}^{-1}} (\theta - T_i) \quad (50)$$

$$Q_o = \frac{1}{S} \frac{R_{so}^{-1} R_2^{-1}}{R_{so}^{-1} + R_2^{-1}} \left( T_o + \frac{1}{R_{so}^{-1}} \alpha SI_{SV} - \theta \right) \quad (51)$$

where $\boldsymbol{\theta} \equiv \boldsymbol{\theta_3}$, $\boldsymbol{R_{si}} \equiv \boldsymbol{R_4}$, $\boldsymbol{R_{so}} \equiv \boldsymbol{R_1}$, $\boldsymbol{q_3} \equiv \boldsymbol{Q_i}$, $\boldsymbol{q_2} \equiv \boldsymbol{Q_o}$, $\boldsymbol{f_1} \equiv \boldsymbol{\alpha S I_{SV}}$ and $\boldsymbol{f_3} \equiv \boldsymbol{0}$.

In Eq. (50)-(51) the heat flux density is defined as positive when comes into the room from outdoor to indoor ambient.

The matrices $\mathbf{A}_S$ and $\mathbf{B}_S$ are the same as those obtained in Section 3.1.1. and a new output matrix $\mathbf{C'}_S$ and a new feed through matrix $\mathbf{D'}_S$ are obtained:

$$\mathbf{C'}_S = \begin{bmatrix} \frac{1}{S}\frac{R_3^{-1}R_{si}^{-1}}{R_3^{-1}+R_{si}^{-1}} \\ -\frac{1}{S}\frac{R_{so}^{-1}R_2^{-1}}{R_{so}^{-1}+R_2^{-1}} \end{bmatrix}, \text{ is the output matrix considering heat flux densities, and}$$

$$\mathbf{D'}_S = \begin{bmatrix} 0 & 0 & 0 & 0 & -\frac{1}{S}\frac{R_3^{-1}R_{si}^{-1}}{R_3^{-1}+R_{si}^{-1}} & 0 & 0 \\ \frac{1}{S}\frac{R_{so}^{-1}R_2^{-1}}{R_{so}^{-1}+R_2^{-1}} & 0 & 0 & 0 & 0 & \frac{R_{so}^{-1}R_2^{-1}}{R_{so}^{-1}+R_2^{-1}}\frac{\alpha}{R_{so}^{-1}} & 0 \end{bmatrix}, \text{ is the feed through matrix}$$

considering heat flux densities.

## 4.2 Bode diagrams of the wall model: inside surface temperature and heat flux density

The transfer functions, related to the inside surface temperature and the indoor heat flux density, can be obtained from Eq. (**45**) which contains the matrices of the state-space representation. For an illustrative first order example, the matrices of the state-space representation are those obtained in Section 4.1; higher order models can be derived by following the same procedure. The Bode plots for a first order model ($N = 3$ nodes) are shown in Figure 10 for the inside surface temperature, $\theta_{si}$, and the inside heat flux density, $Q_i$, as a function the inside air temperature, $T_i$, the outside air temperature, $T_o$ and the solar irradiance $I_s$. The transfer functions considering as inputs $T_o$ and $I_s$ are strictly proper and the output signals, $\theta_{si}$ and $Q_i$, tend to be totally damped for high frequencies after the cut-off (red dashed line), i.e. the wall acts as a low pass filter. The cut-off of a wall is given by its thermal resistance and its heat capacity (Naveros & Ghiaus, 2015). Nonetheless, the transfer functions considering the input $T_i$ are proper. Hence the output signal does not tend to be totally damped to high frequencies (Figure 10c and d); the transfer function of the inside surface temperature acts as a low pass filter (Figure 10c), but it is found that the transfer function of the inside heat flux density acts as a high pass filter against physical experience, i.e. high frequencies are more amplified than low frequencies (Figure 10d).

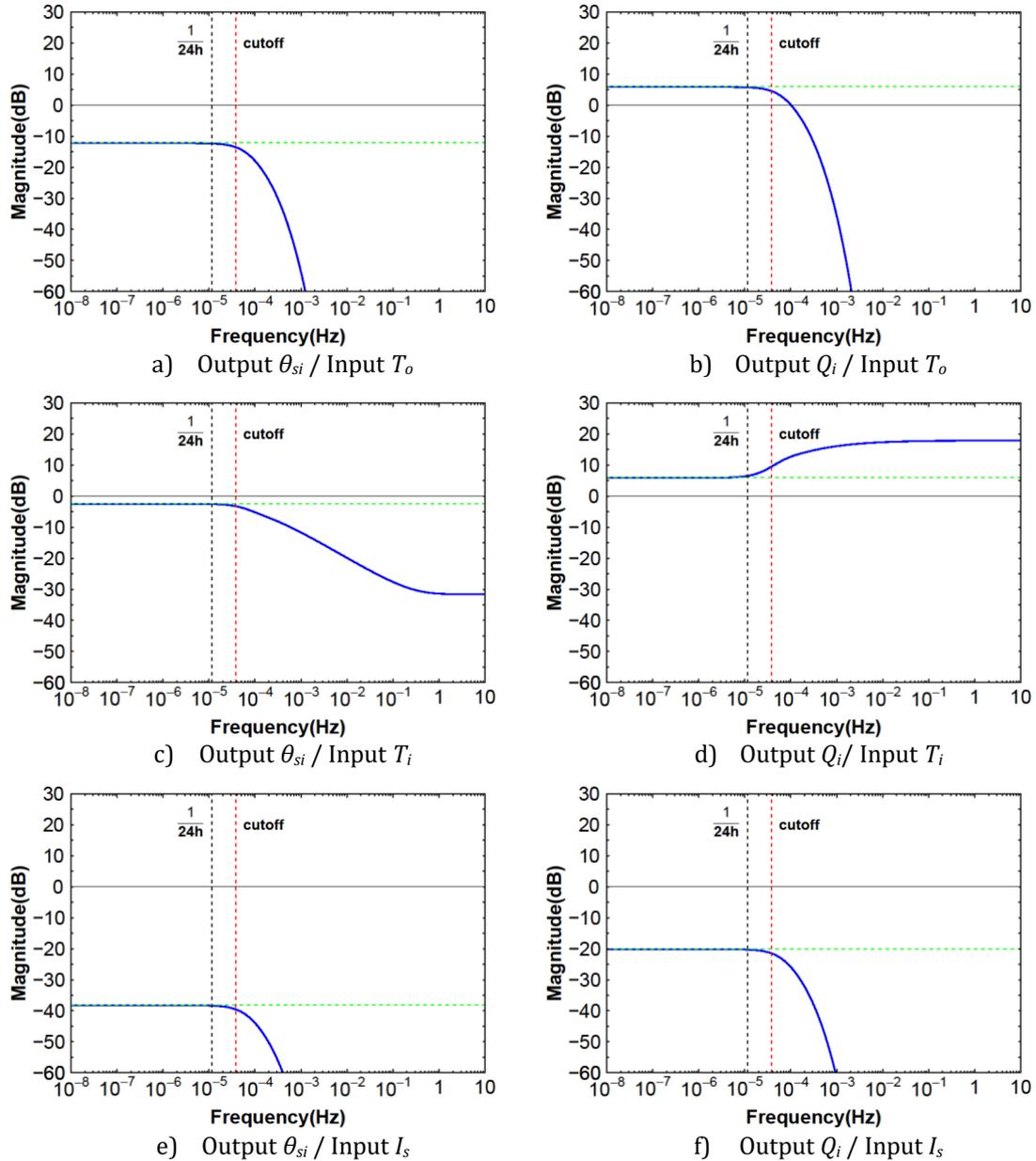

**Figure 10.** Bode magnitude plots for a wall considering the inside surface temperature, $\theta_{si}$, and the inside heat flux density, $Q_i$.

In the case of considering the outside surface temperature and the outside heat flux density regarding as outputs and outdoor ambient temperature and solar irradiance as inputs, the transfer functions obtained are also proper, not strictly proper, **Error! Reference source not found.**a, b, e and f. The problem of obtaining transfer functions acting as high pass filters, also appears for transfer functions representing the heat flux density only (**Error! Reference source not found.**b and f).

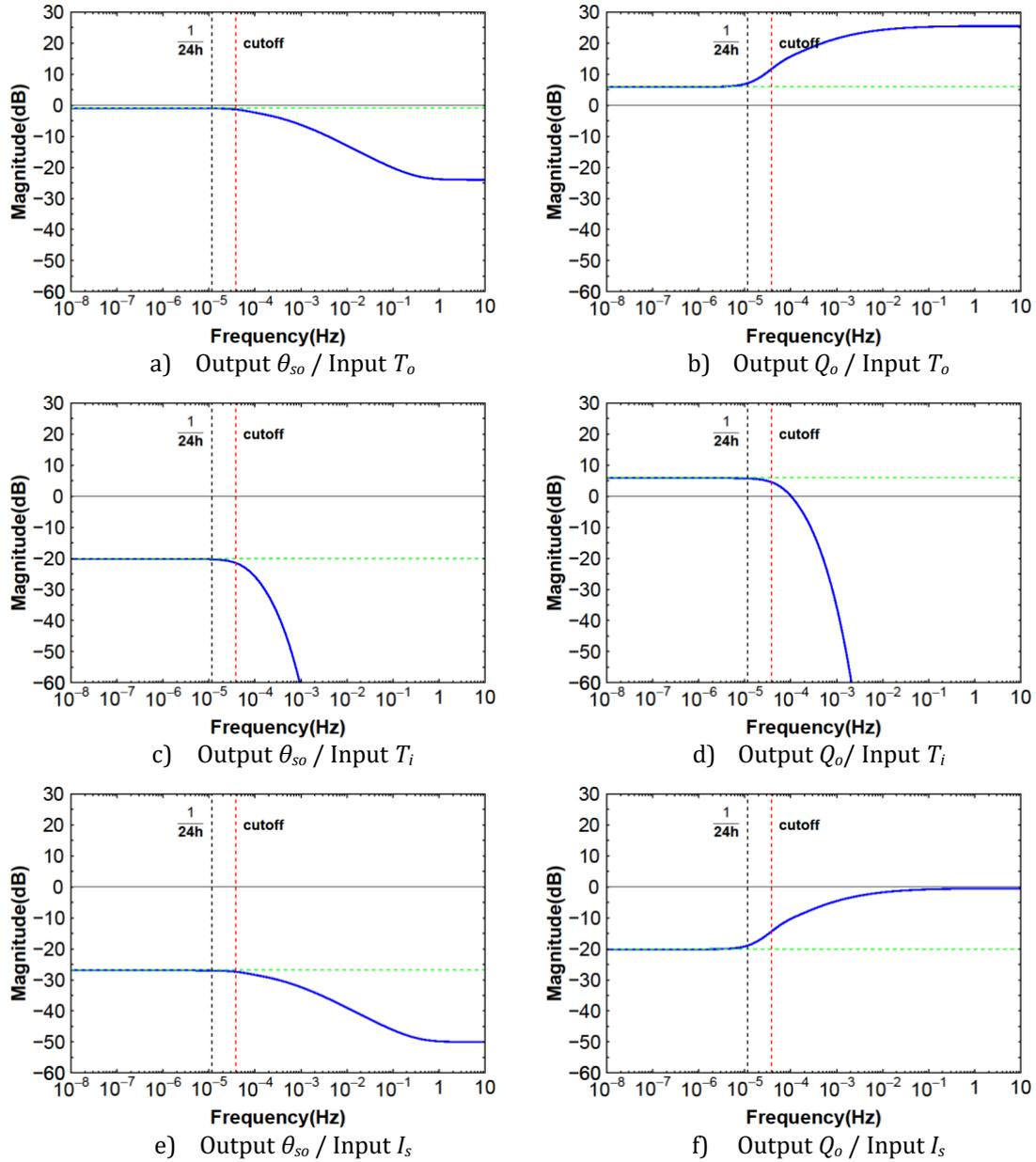

**Figure 11.** Bode magnitude plots for a wall considering the outside surface temperature, $\theta_{so}$, and the inside heat flux density, $Q_o$.

The values of the thermal resistances and the thermal capacitance of the wall used for simulating the transfer functions represented in Figure 10 and **Error! Reference source not found.**, are: $R_{wall}^{-1} = \sum_{i=2}^{N} R_i = 20$ W/K, $R_{so}^{-1} = 200$ W/K, $R_{si}^{-1} = 80$ W/K, $C_{wall} = \sum_{i=2}^{N-1} C_i = 2.1 \times 10^6$ J/K, $S = 10$m² and $\alpha = 1$. The term related to the absorptivity is assumed one by the sake of simplicity, since this term only introduces a constant reduction to the input which is equal for all the frequencies present in the solar irradiance signal (it is important to note that frequencies are related to the variation of the signal in time not to the wavelength of the sun light).

# 5  Discussion

By doing a frequency analysis, it has been proved that the use of heat flux densities and surface temperatures implies the use of not strictly proper transfer functions instead of strictly proper transfer functions. As heat transfer is a dissipative process, it is expected to obtain a transfer function relating the input signal to the output signal which acts as a low pass filter. Nonetheless, it has been shown that not strictly proper transfer functions related to heat flux densities act as high pass filters, which does not occur in the case of transfer functions related to surface temperatures.

This theoretical issue should be taken into account when heat flux measurements are used for determining wall thermal performances. Hence heat flux densities acting as outputs that have a proper, but not strictly proper, transfer function would be theoretically limited to be used when high frequencies are present in input signals. The limitation could be significant for driven forces which presents high frequencies in their measured signals. For instance, this theoretical issue mainly implies a limitation in addition to experimental source errors when heat flux meters are exposed to inputs as solar irradiance Figure 12, which is an input signal that may content high frequencies components higher than the cut-off frequency (Naveros & Ghiaus, 2015)). In the case of the wall simulated on Figure 10 and **Error! Reference source not found.** the cut-off begins later than the frequency corresponding to 1/12h, which is the second main component of the solar irradiance besides the main frequency component corresponding to the inverse of the daily cycle (1/24h), Figure 12.

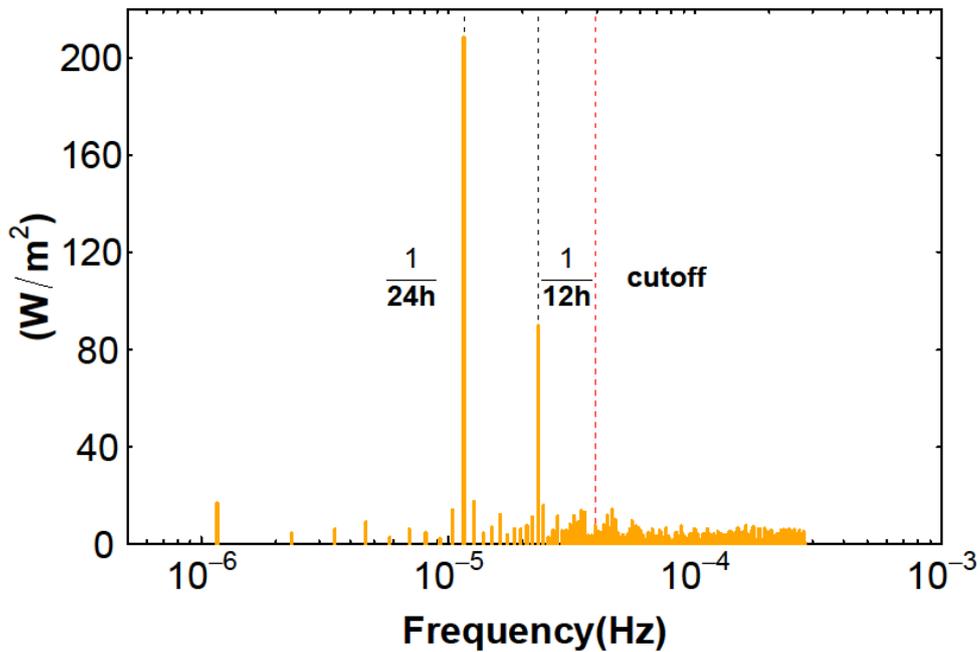

Figure 12. Solar irradiance spectrum (Naveros & Ghiaus, 2015).

In such a case, as shown in **Error! Reference source not found.**f, the outside heat flux density will reduce the component signals corresponding to high frequencies less than those corresponding to low frequencies and this will imply to have a high pass filter instead of a low pass filter. This is due to the fact that the measurements of the heat flux meters are usually based on the algebraic equation: $q_{hfm} = k_{hfm}\Delta T_{hfm}$ (Kaviany, 2013). This theoretical deduction may be observed in measurements of heat flux densities given by heat flux meters placed outdoors (Figure **13**). Particularly, a comparison between Figure 12 and Figure **13** shows that the rate between amplitudes at different frequencies (amplitude at 1/12h divided by amplitude at 1/24h) is greater for the outside heat flux density, the rate is 0.53, than for the solar irradiance, the rate is 0.43; i.e. it is observed in the measurements of the heat flux meter that the amplitude corresponding to the 1/12h frequency is less damped than the amplitude corresponding to the 1/24h frequency.

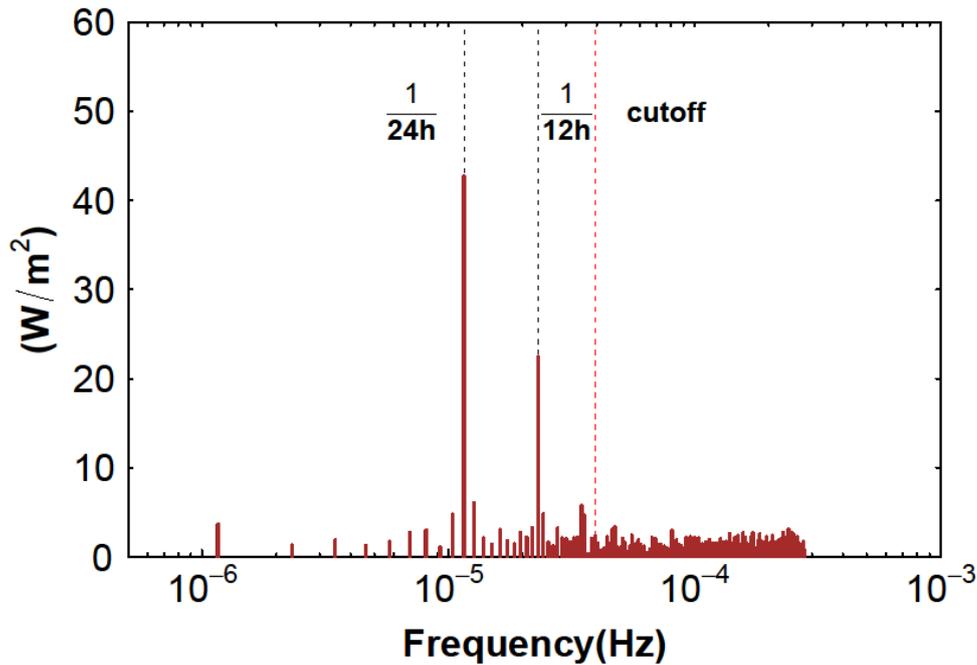

Figure 13. Outside heat flux density spectrum (Naveros & Ghiaus, 2015).

On the other hand, the use of surface temperature measurements would imply the use of low pass filters as expected in a dissipative process. The measurement of surface temperatures by using thermocouples would present a similar limitation regarding to their locations than the use of heat flux meters, i.e. using thermocouples or heat flux meters would require the use of sensor arrays at different locations for dealing with thermal performance of walls (Meng, et al., 2015). Meng et al. deal with different errors sources related to the use of thermocouples and heat flux meters and they fix their attention on sensor locations. Nonetheless, an alternative to be considered in future studies, which would imply an advantage to the use of heat flux meters or thermocouple, would be the use of thermal cameras which would give a measure of surface temperatures at different locations limited by their image resolution power only.

Lastly, it should be noted the known limitation of classical theories, as heat transfer theory, due to the assumption of forces which can act instantaneously at a distance (Roubicek & Valasek, 2002). The problem of forces acting instantaneously at a distance resides in the use of algebraic equations. In fact, algebraic equations are used for facilitating the resolution of the problem of a system of ordinary differential equations (ODE) by using a system of differential and algebraic equations (DAE). In this case, the transformation of the heat equation into a system of DAE is a simplification which implicitly assumes the existence of forces which act instantaneously at a distance. This is not a problem since the use of the classical heat transfer theory is useful at buildings scale and it is a well-known limitation of classic theories, but it remains hidden if it is not explicitly noted.

# 6   Conclusions

The transfer of a signal from input to output implies a change of amplitude as a function of frequency. Since any heat transfer process is dissipative, the amplitudes of the output signal needs to be more damped at high frequencies than at low frequencies. In the case of surface temperatures acting as outputs, this fact is observed. Nonetheless, in the case of heat flux densities acting as outputs, though the heat flux density is limited by the resistances of the system, it is observed that amplitudes at high frequencies are less damped than amplitudes at low frequencies.

For this reason, there are two main problems in the use of heat flux meters for wall thermal performances: an experimental problem due to dynamic conditions of in-situ experiments, and another theoretical problem due to the assumption of the classical theory about the existence of forces acting instantaneously at a distance. The assumption about the existence of instantaneous forces is the basis for the use of algebraic equations in the heat flow method. In addition to the experimental source errors, such an assumption limits the use of heat flux meters when the assessment of wall thermal performances is done under real weather conditions.

The theoretical problem is highlighted in frequency domain since proper transfer functions acting as high pass filters, against physical experience, may be obtained when heat flux densities are considered as outputs. The theory is not valid at that point and cannot be used for explaining the behaviour at every high frequency dependent on input signals. Then, the domain of validity of models using the heat flux meter method is limited by this fact for signal inputs with frequencies close to the cut-off frequency of the thermal system or higher than it. Therefore, the use of heat flux meters for measuring heat flux densities through walls should be mainly avoided outdoors even if heat flux meters are protected from solar irradiance. Furthermore, the limitation would be greater for heavy walls, which will have cut-off frequencies at lower frequencies than light walls.

## ACKNOWLEDGEMENTS

Funding for this project was provided by a grant from "Région Rhône-Alpes", by the Spanish Ministry of Economy and Competitiveness within the program Juan de la Cierva-Formación and under the project TIN2015-64776-C3-1-R.

# REFERENCES

Bacher, P. & Madsen, H., 2011. Identifying suitable models for the heat dynamics of buildings. *Energy and Buildings,* 43(7), pp. 1511-1522.

Biddulph, P. et al., 2014. Inferring the thermal resistance and effective thermal mass of a wall using frequent temperature and heat flux measurements. *Energy and Buildings,* 78(0), pp. 10-16.

Bloem, J. J. et al., 1994. *Workshop on Application of System Identification in Energy Savings in Buildings: October 25-27, 1993.* s.l.:Joint Research Centre, Commission of the European Communities.

Carslaw, H. & Jaeger, J., 1986. *Conduction of Heat in Solids.* s.l.:Clarendo Press.

Chen, Y., Athienitis, A. K. & Galal, K. E., 2013. Frequency domain and finite difference modeling of ventilated concrete slabs and comparison with field measurements: Part 1, modeling methodology. *International Journal of Heat and Mass Transfer,* 66(0), pp. 948-956.

Cipriano, J. et al., 2016. Development of a dynamic model for natural ventilated photovoltaic components and of a data driven approach to validate and identify the model parameters. *Solar Energy,* Volume 129, pp. 310-331.

de Wilde, P., 2014. The gap between predicted and measured energy performance of buildings: A framework for investigation. *Automation in Construction,* 41(0), pp. 40-49.

EN-ISO-9869, 1994. *Thermal Insulation - Building Elements - In-situ Measurement of Thermal Resistance and Thermal Transmittance.* s.l.:EN ISO 9869.

Foucquier, A. et al., 2013. State of the art in building modelling and energy performances prediction: A review. *Renewable and Sustainable Energy Reviews,* Volume 23, pp. 272-288.

Fourier, J. B. J., 1822. *Théorie analytique de la chaleur.* s.l.:s.n.

Ghiaus, C., 2013. Causality issue in the heat balance method for calculating the design heating and cooling load. *Energy,* 50(0), pp. 292-301.

Incropera, F. P., 2006. *Fundamentals of Heat and Mass Transfer.* s.l.:John Wiley \& Sons.

Jimenez, M. J. & Heras, M. R., 2005. Application of multi-output ARX models for estimation of the U and g values of building components in outdoor testing. *Solar Energy,* 79(3), pp. 302-310.

Jimenez, M. J., Porcar, B. & Heras, M. R., 2008. Estimation of building component UA and gA from outdoor tests in warm and moderate weather conditions. *Solar Energy,* 82(7), pp. 573-587.

Kaviany, M., 2013. Essentials of Heat Transfer: Principles, Materials, and Applications. *Contemporary Physics,* 54(3), pp. 173-174.

Martin, K. et al., 2012. Equivalent wall method for dynamic characterisation of thermal bridges. *Energy and Buildings,* Volume 55, pp. 704-714.

Meng, X. et al., 2015. Factors affecting the in situ measurement accuracy of the wall heat transfer coefficient using the heat flow meter method. *Energy and Buildings,* Volume 86, pp. 754-765.

Milman, M. & Petrick, W., 2000. A note on the solution to a common thermal network problem encountered in heat-transfer analysis of spacecraft. *Applied Mathematical Modelling,* 24(12), pp. 861-879.

Narasimhan, T. N., 1999. Fourier's heat conduction equation: History, influence, and connections. *Reviews of Geophysics,* 37(1), pp. 151-172.

Naveros, I. et al., 2014. Setting up and validating a complex model for a simple homogeneous wall. *Energy and Buildings,* Volume 70, pp. 303-317.

Naveros, I. & Ghiaus, C., 2015. Order selection of thermal models by frequency analysis of measurements for building energy efficiency estimation. *Applied Energy,* 139(0), pp. 230-244.

Naveros, I., Ghiaus, C., Ruíz, D. & Castaño, S., 2015. Physical parameters identification of walls using ARX models obtained by deduction. *Energy and Buildings,* Volume 108, pp. 317-329.

Naveros, I., Jiménez, M. J. & Heras, M. R., 2012. Analysis of capabilities and limitations of the regression method based in averages, applied to the estimation of the U value of building component tested in Mediterranean weather. *Energy and Buildings,* Volume 55, pp. 854-872.

Rabl, A., 1988. Parameter Estimation in Buildings: Methods for Dynamic Analysis of Measured Energy Use. *Journal of Solar Energy Engineering,* 110(1), pp. 52-66.

Roubicek, T. & Valasek, M., 2002. Optimal control of causal differential-algebraic systems. *Journal of Mathematical Analysis and Applications,* 269(2), pp. 616-641.

Strachan, P. & Baker, P., 2008. Outdoor testing, analysis and modelling of building components. *Building and Environment,* 43(2), pp. 127-128.

Strang, G., 1986. *Introduction to Applied Mathematics.* s.l.:Wellesley-Cambridge Press.

Strang, G., 2007. *Computational Science and Engineering.* s.l.:Wellesley-Cambridge Press.

Vieites, E., Vassileva, I. & Arias, J. E., 2015. European Initiatives Towards Improving the Energy Efficiency in Existing and Historic Buildings. *Energy Procedia,* Volume 75, pp. 1679-1685.